# A Methodology for Thermal Simulation of Interconnects Enabled by Model Reduction with Material Property Variation

Wangkun Jia and Ming-C. Cheng[*]
Department of Electrical and Computer Engineering
Clarkson University, Potsdam, NY 13699-5720
[*]Corresponding author: mcheng@clarkson.edu

**Abstract**

A thermal simulation methodology is developed for interconnects enabled by a data-driven learning algorithm accounting for variations of material properties, heat sources and boundary conditions (BCs). The methodology is based on the concepts of model order reduction and domain decomposition to construct a multi-block approach. A generic block model is built to represent a group of interconnect blocks that are used to wire standard cells in the integrated circuits (ICs). The blocks in this group possess identical geometry with various metal/via routings. The data-driven model reduction method is thus applied to *learn* material property variations induced by different metal/via routings in the blocks, in addition to the variations of heat sources and BCs. The approach is investigated in two very different settings. It is first applied to thermal simulation of a single interconnect block with similar BCs to those in the training of the generic block. It is then implemented in multi-block thermal simulation of a FinFET IC, where the interconnect structure is partitioned into several blocks each modeled by the generic block model. Accuracy of the generic block model is examined in terms of the metal/via routings, BCs and thermal discontinuities at the block interfaces.

## 1. Introduction

Temperature escalation in integrated circuits (ICs) due to Joule heating has emerged as one of the most critical issues in integrated circuit (IC) design for decades [1]-[11]. This results from the rapid increase in the device and interconnect densities in semiconductor chips. High thermal gradients and hot spots induced by Joule heating significantly degrade performance and reliability of ICs in various IC technologies [1]–[9]. In addition to self-heating effects in devices [12]–[16], high temperature in interconnects also imposes serious problems. For example, metal resistance and signal delays increase as temperature rises, which limits the operating speed of ICs [8], [9], [17]. Moreover, the failure rate of interconnects due to electromigration is accelerated by high temperature [17]–[19]. It is thus essential to predict accurate thermal distributions over the interconnects in IC design.

Simulations for predicting temperature distributions in ICs from the gate to the system level have usually been based on efficient lumped thermal element or compact thermal models [20]–[28]. These approaches rely on *RC* thermal elements and/or averaging concepts to minimize the computing time. They are however limited to regional average temperature and not able to offer accurate temperature gradients or hot spots in semiconductor chips. To provide more detailed and accurate thermal profiles, direct numerical simulation (DNS) is needed but its intensive computational demand is prohibitive for thermal simulation of large-scale semiconductor chips.

In order to offer both efficiency and accuracy of the thermal prediction in ICs, alternatives based on the projection-based reduced-order models have been applied in recent years [29]–[36]. Many of these approaches utilize proper orthogonal decomposition (POD) [29]–[36]. The POD methods have been investigated in many different fields including fluid dynamics [37]–[40], micro-electro-mechanical systems (MEMS) [41]–[42], heat transfer [29]–[36], etc. The POD generates orthogonal basis functions from solution data of the DNSs in a domain subjected to the parametric variations that for heat transfer problems usually include spatial/temporal power sources and boundary conditions (BCs). The decomposition optimizes the basis functions (or POD modes) specifically tailored to the geometry and parametric variations of the problem using a data-driven learning process. With the heat conduction equation projected to the POD modes, the approach is able to significantly reduce the numerical degree of freedom (DoF) needed to accurately predict the thermal profile in the domain.

The major drawback of the POD method is the time-consuming process for data collection from DNSs, generation of POD modes and calculations of POD model parameters for a large domain with a fine resolution. One may also question the usefulness of the POD method for engineering applications since this "training" or "learning" process for the POD modes requires full-scale DNSs with enough variations of heat sources and BCs. It becomes prohibited in a very large domain with a fine spatial resolution, especially in a dynamic problem.

To resolve this issue, a multi-block POD methodology was proposed in our previous work [31], [33], which implements the concept of reduced basis elements [43], [44] in POD, similar to the concept based on space vector clustering [45]. The multi-block methodology implements domain decomposition in the POD method, which partitions a large domain into smaller building blocks. Each of these blocks is projected onto a functional space described by its POD modes. This approach offers several advantages. First, computational resources needed in DNSs to collect thermal data to generate POD modes and parameters increase exponentially with the size of the domain. With smaller blocks, the POD modes and parameters can be generated more efficiently. To construct a large domain, the projected building blocks can be *glued* together with the interface continuity enforced by the interior penalty discontinuous Galerkin (DG) method [46], [47]. Second, domain decomposition possesses a nature advantage of parallel computing and has been proven effective in parallel/distributed computing settings [48]–[51]. Finally, IC design at all levels always utilizes identical repeating functional circuit blocks, such as standard cells, caches, memory units, CPU/GPU cores, GPU streaming multiprocessors, etc. The POD modes and parameters of these generic building blocks can be generated and stored in a library for each design level, which can then offer cost-effective thermal simulation and thermal management of large IC structures.

Most POD applications to thermal simulations have been based on single-block approaches [29], [30], [34]–[37], the multi-block POD approach with generic blocks have recently been successfully applied to device and IC structures [31], [33]. It was shown [31] that the multi-block POD model is able to predict accurate 3D dynamic thermal distributions and small-size hot spots in a FinFET IC structure subjected to random power pulses induced by digital input voltages. It was demonstrated that the multi-block POD simulation offers a reduction in DoF by 5 to 6 orders of magnitude with high accuracy compared to DNS.

In [31], the interconnects were not included in the multi-block POD simulation because it is difficult to find interconnect building blocks that are repeatedly used in different locations of ICs. Metal routings in ICs consist of a wide range of variations, which are selected based on the placement of the functional blocks. However, within a specific group of interconnects with similar features, it is still possible to select a generic block to develop a multi-block POD approach for interconnect thermal simulation if the material property variation (MPV) is considered in the POD-mode training. In the proposed multi-block MPV-POD methodology, the POD modes in each selected block are then trained to experience variation of thermal properties between the metal and dielectric at the locations where the metal lines and vias appear. To construct a multi-block MPV-POD thermal model, several small interconnect blocks with various metal/via routings modeled by the same generic-block POD model can then be glued together. To the best of our knowledge, this is the first study considering MPV in POD. It should be noted that, unlike the standard cells or many other functional circuit blocks, it is difficult to select building blocks that can be used in different groups of interconnects in ICs. Different types of generic blocks will be needed for different interconnect configurations. In this work, the developed MPV-POD methodology is applied to the interconnects that are used to wire standard cells. With the demonstration of a simple interconnect configuration to examine the concept and effectiveness of the MPV-POD approach, this study perhaps paves the way toward developing a more sophisticate multi-block MPV-POD methodology for thermal simulation of on-chip and off-chip interconnects.

After an overview of the conventional POD method in Section 2 for thermal problems, the MPV-POD methodology is presented in Section 3. The MPV-POD approach for interconnect thermal simulation is illustrated in Section 4 step-by-step including the settings of the simulation domains for data collection and the generation of POD eigenvalues and modes. In Section 5, the MPV-POD method is demonstrated in a single-block interconnect structure and a small multi-block IC structure, compared against the DNSs. Discussions of the findings are given in Section 6, and conclusions are finally presented in Section 7.

## 2. Overview of POD Thermal Simulation Methodology
### 2.1. POD for a single-block domain

POD generates a set of modes from spatial/temporal thermal data accounting for the parametric variation. This is done by seeking a POD mode $\varphi(\vec{x})$ that maximizes its mean square inner product with an ensemble thermal data $T(\vec{x},t)$ in the domain $\Omega$,

$$\left\langle \left( \int_\Omega T(\vec{x},t)\varphi\, d\Omega \right)^2 \right\rangle \Big/ \int_\Omega \varphi^2 d\Omega, \tag{1}$$

where $\langle \cdot \rangle$ indicates an average over ensemble data sets observed in time in our study. This can also be done in static problems where the observations are carried out in response to a range of heat source strengths and BCs. This maximization process ensures that the component projected onto the POD mode contains the maximum least squares (LS) information of the thermal behavior described by the thermal data [52], [53]. In the space orthogonal to this mode, the process can be performed again to generate the second mode. Repetition of the process in this fashion results in an orthogonal set of POD modes.

Applying the variational calculus to (1), this problem can be reformulated to the Fredholm equation of the second kind,

$$\int_{\vec{x}'} R(\vec{x},\vec{x}')\varphi(\vec{x}')\, d\vec{x}' = \lambda \varphi(\vec{x}), \tag{2}$$

where $R(\vec{x},\vec{x}')$ is an two-point correlation tensor given by

$$R(\vec{x}, \vec{x}') = \langle T(\vec{x}, t) \otimes T(\vec{x}', t) \rangle, \tag{3}$$

with $\otimes$ as the tensor product, and $\lambda$ is the POD eigenvalue of $R$ and represents the mean squared temperature captured by the corresponding POD mode. This decomposition process leads to an eigenvalue problem represented by (2) for $R$. Once the POD modes are found, temperature $T(\vec{x}, t)$ can be represented by a linear combination of the POD modes,

$$T(\vec{x}, t) = \sum_{i=1}^{M} a_i(t) \varphi_i(\vec{x}), \tag{4}$$

where $M$ is the selected number of modes or DoF for the temperature solution and $a_i$ is the time dependent coefficient for each mode. The dimension of the eigenvalue problem given in (2) for a large-scale multi-dimensional structure may be enormously large and numerically prohibitive. To generate the POD modes and eigenvalues more efficiently, the method of snapshots [29], [31], [54] is applied to convert the eigenvalue problem from a space domain to a sampling domain whose dimension is determined by the smaller number of samples/snapshots. A brief overview of the method of snapshots is presented in Appendix.

To predict the temperature in (4), a set of equations for $a_j$ is generated by projecting the heat conduction equation onto an eigenspace using the Galerkin projection method,

$$\int_\Omega \left( \varphi \frac{\partial \rho C T}{\partial t} + \nabla \varphi \cdot k \nabla T \right) d\Omega = \int_\Omega \varphi P_d d\Omega - \int_\Gamma \varphi(-k\nabla T \cdot \vec{n}) d\Gamma, \tag{5}$$

where $k$ is the thermal conductivity, $P_d$ the power density, $\rho$ the density, $C$ the specific heat, $\Gamma$ the boundary surface and $\vec{n}$ the outward normal vector of the surface. With a selected number of modes $M$, the spatial integrals in (5) can be pre-evaluated to construct a set of $M$ ordinary differential equation for $a_i$,

$$\sum_{j=1}^{M} c_{i,j} \frac{da_j}{dt} + \sum_{j=1}^{M} g_{i,j} a_j = P_{pod,i}, \quad i = 1 \text{ to } M, \tag{6}$$

where $c_{i,j}$ and $g_{i,j}$ are elements of the thermal capacitance and conductance matrices in the POD space and given by

$$c_{i,j} = \int_\Omega \rho C \varphi_i \varphi_j d\Omega \quad \text{and} \quad g_{i,j} = \int_\Omega k \nabla \varphi_i \cdot \nabla \varphi_j d\Omega, \tag{7}$$

and $P_{pod,i}$ represents the projected power density in $\Omega$ and heat flux across $\Gamma$ along the $i$th POD mode,

$$P_{pod,i} = \int_\Omega \varphi_i P_d(\vec{x}, t) d\Omega - \int_\Gamma \varphi_i (-k\nabla T \cdot \vec{n}) d\Gamma. \tag{8}$$

Power density $P_d$ in metal is induced by Joule heating,

$$P_d = \vec{J} \cdot \vec{E} = J^2 / \sigma. \tag{9}$$

where $\vec{J}$ is the current density, $\vec{E}$ the electric field, and $\sigma$ the copper conductivity. To obtain realistic current density in data collections and demonstrations, circuit simulations of FinFET ICs, such as the FinFET circuit shown in Fig. 1, including the interconnect routings with metal resistance, are performed in Spice. Based on the current in each metal wire obtained from Spice simulation, together with the metal wire cross section and conductivity, $J$ and $P_d$ are calculated. With the determined $a_i$, the temperature can be predicted from (1). The off-diagonal terms of $c_{i,j}$ are usually small and ignored in previous studies [29]-[31]. In this study, they are all included.

### 2.2. *POD formulation for multi-block structure*

When several blocks adjoin together, thermal continuity at the interface needs to be appropriately enforced in the last term of (5) or the second term of (8). The interior penalty DG method [46], [47] is applied to the interface and (5) becomes

$$\int_\Omega \left(\varphi \frac{\partial \rho C T}{\partial t} + \nabla\varphi \cdot k\nabla T\right) d\Omega - k\int_\Gamma \left([\![T]\!]\cdot\{\nabla\varphi\} + \{\nabla T\}\cdot[\![\varphi]\!]\right) d\Gamma + k\int_\Gamma \mu [\![T]\!]\cdot[\![\varphi]\!] d\Gamma = \int_\Omega \varphi P_d(x,t) d\Omega, \quad (10)$$

where $\mu$ is a penalty constant defined as $N_\mu/dx$ ($dx$ is the local element size and $N_\mu$ as the non-unit penalty number), and $\{\cdot\}$ and $[\![\cdot]\!]$ are the average and difference across the interface, respectively (see [31]). A large positive value of $\mu$ is usually needed to stabilize the numerical result. Numerical accuracy and stability influenced by $N_\mu$ will be examined in Section 5.2.

For a domain consisting of $N$ projected blocks, (10) reduces to an $N$-block POD model given as

$$\begin{bmatrix} \mathbf{C}_1 & 0 & \cdots & 0 \\ 0 & \mathbf{C}_2 & \cdots & 0 \\ \vdots & \vdots & \ddots & \vdots \\ 0 & 0 & \cdots & \mathbf{C}_N \end{bmatrix} \frac{d}{dt}\begin{bmatrix} \vec{a}_1 \\ \vec{a}_2 \\ \vdots \\ \vec{a}_N \end{bmatrix} + \begin{bmatrix} \mathbf{G}_1 + \sum_{q=2}^{N}\mathbf{G}_{1,B_{1,q}} & \mathbf{G}_{1,2} & \cdots & \mathbf{G}_{1,N} \\ \mathbf{G}_{2,1} & \mathbf{G}_2 + \sum_{q=1,q\neq 2}^{N}\mathbf{G}_{2,B_{2,q}} & \cdots & \mathbf{G}_{2,N} \\ \vdots & \vdots & \ddots & \vdots \\ \mathbf{G}_{N,1} & \mathbf{G}_{N,2} & \cdots & \mathbf{G}_N + \sum_{q=1}^{N-1}\mathbf{G}_{N,B_{N,q}} \end{bmatrix} \cdot \begin{bmatrix} \vec{a}_1 \\ \vec{a}_2 \\ \vdots \\ \vec{a}_N \end{bmatrix} = \begin{bmatrix} \vec{P}_1 \\ \vec{P}_2 \\ \vdots \\ \vec{P}_N \end{bmatrix}, \quad (11)$$

where the subscript $B_{p,q}$ denotes the boundary surface in the $p$th block between the $p$th and $q$th blocks, $\mathbf{C}_p$ and $\mathbf{G}_p$ are the $c_{i,j}$ and $g_{i,j}$ matrices of the $p$th block given in (7), $\mathbf{G}_{p,B_{p,q}}$ is given as

$$\mathbf{G}_{p,B_{p,q}} = \begin{bmatrix} g_{p,B_{p,q},1,1} & g_{p,B_{p,q},1,2} & \cdots & g_{p,B_{p,q},1,M_p} \\ g_{p,B_{p,q},2,1} & g_{p,B_{p,q},2,2} & \cdots & g_{p,B_{p,q},2,M_p} \\ \vdots & \vdots & \ddots & \vdots \\ g_{p,B_{p,q},M_p,1} & g_{p,B_{p,q},M_p,2} & \cdots & g_{p,B_{p,q},M_p,M_p} \end{bmatrix}, \quad (12)$$

and $\mathbf{G}_{p,g}$ is given as

$$\mathbf{G}_{p,q} = \begin{bmatrix} g_{p,q;1,1} & g_{p,q;1,2} & \cdots & g_{p,q;1,M_q} \\ g_{p,q;2,1} & g_{p,q;2,2} & \cdots & g_{p,q;2,M_q} \\ \vdots & \vdots & \ddots & \vdots \\ g_{p,q;M_p,1} & g_{p,q;M_p,2} & \cdots & g_{p,q;M_p,M_q} \end{bmatrix} \quad (13)$$

If the $p$th and $q$th do not adjoin each other, $\mathbf{G}_{p,B_{p,q}}=0$, and $\mathbf{G}_{p,g}=0$. The thermal coupling conductance matrices in (12) and (13) is given in the following equations respectively,

$$g_{p,B_{p,q},i,j} = -k\int_{\Gamma_p}\left(\frac{1}{2}\varphi_{p,j}\nabla\varphi_{p,i} + \frac{1}{2}\varphi_{p,i}\nabla\varphi_{p,j} - \mu\varphi_{p,i}\varphi_{p,j}\right) d\Gamma \quad (14)$$

$$g_{p,q;i,j} = -k\int_\Gamma \left(-\frac{1}{2}\varphi_{q,j}\nabla\varphi_{p,i} + \frac{1}{2}\varphi_{p,i}\nabla\varphi_{q,j} + \mu\varphi_{p,i}\varphi_{q,j}\right) d\Gamma. \quad (15)$$

With a large number of blocks in a structure, the block $\mathbf{G}$ matrix in (11) becomes sparse. The nonzero blocks in a row are neighbored by other elements.

## 3. POD with Material Property Variation

In this study, we extend the POD thermal simulation methodology to account for the MPV between metal and dielectric (oxide). More specifically, for a selected interconnect block that may include one or more layers of metal and interlayer dielectric (ILD), the POD modes are trained to include the effects of the variation when a metal line or a via appears in the dielectric. Thermal solution data collected from DNSs to generate/train the POD modes must include the MPV in the selected block to offer influences of the MPV on the solution of (6). It is however difficult for POD modes to effectively capture the effects of the MPV if the metal lines and/or vias appear at arbitrary locations. For such a case, a large number of POD modes will be needed to predict accurate thermal solution; namely the DoF will not be significantly reduced. It is therefore more effective to train a finite set of POD modes for a specific group of interconnects that have similar metal/via routing features to reveal some generic blocks. Each generic block is designated for blocks with the same geometric shape and identical dimensions, consisting of one or more metal and ILD layers. Each generic block can be trained to generate POD modes that are able to account for effects of MPV induced by various metal/via routings.

There may be several different generic interconnect blocks (e.g., with different block dimensions or with wider metal lines or larger pitches) needed to cover different groups of interconnects. Using this approach, once each selected generic block is trained and projected onto its POD space, their modes and model parameters can be stored in a technology library. To build a large interconnect structure, several projected generic blocks can be glued together to construct a multi-block MPV-POD thermal model, as presented in Section 2.2. There may also be a handful of nonstandard interconnect blocks that do not belong to any of the generic blocks. The POD modes for these individual blocks can also be generated either with or without MPV and implemented in the multi-block structure.

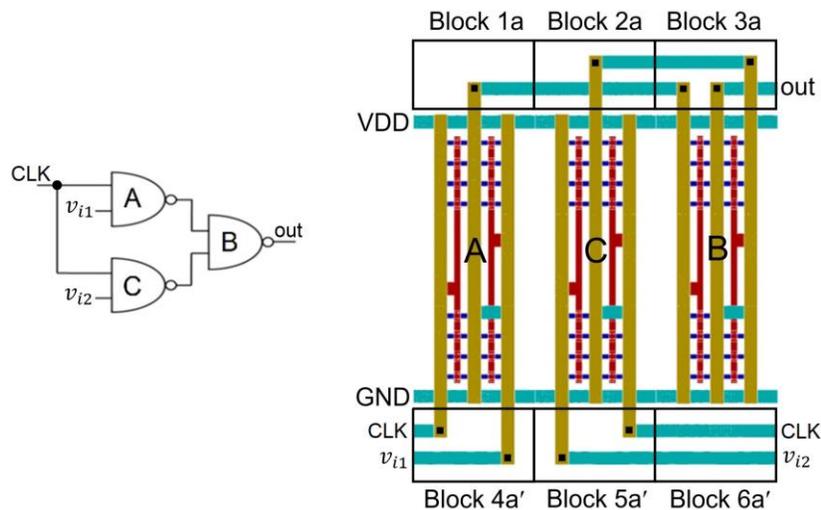

Fig. 1. Circuit and layout for a FinFET IC with horizontal M1 in blue, vertical M2 in yellow, poly in red and vias in black. The NAND-gate structures (labeled as A, B and C) are identical and given in Fig. 2.

A simple FinFET IC shown in Fig. 1 is used to elucidate the concept of the MPV-POD approach. The IC consists of 3 identical FinFET NAND2 gates with the NAND2 structure given in Fig. 2. For ICs using standard cells shown in Fig. 1, the interconnects reveal similar features that can be utilized to identify a generic block. As illustrated in Fig. 1, the interconnects are partitioned into 6 blocks, each with the same shape and identical dimensions. In this case, each block includes Metal-1 (M1), Metal-2 (M2), ILDs and the substrate, and one generic block shown in Fig. 3 can be used to represent a group of interconnect blocks with all possible metal/via routings specified in Fig. 3 based on the technology design rules. The generic block includes 6 possible vias and 14 possible metal segments with square metal pieces connecting the neighboring segments and vias. Each of the 6 interconnect blocks with a specific metal/via routing in Fig. 1 belongs to this generic block or its mirror symmetric block. These 6 blocks or their mirror symmetric blocks are redrawn in Fig. 4 with more clearly defined metal segments. For example, Block 2a consists of M1 Segments 5-8 and M2 Segments 10 and 13 connected to M1 Segments 3 and 4 by a via. Also, Block 4a, the mirror symmetric block of Block 4a′ in Fig. 1, includes M2 Segment 12 connected to M1 Segment 5 by a via and M2 Segments 11 and 14 connected to M1 Segments 1-3 by a via.

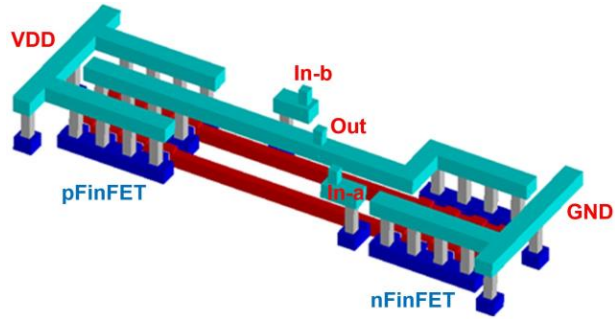

Fig. 2. FinFET NAND2 structure with 4 fins in each FinFET.

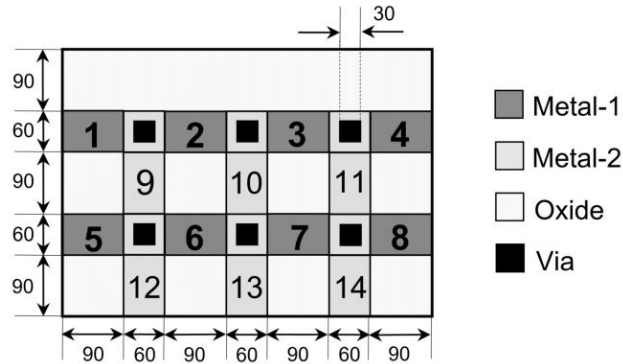

Fig. 3. A generic interconnect block with all possible metal lines and vias whose sizes are labeled in nm. M1 and M2 thicknessess are all equal to 60 nm. Dielectric thickness between two metal layers is 100 nm. Each of the metal segment is labled by a number.

    The method of snapshots [29], [31], [54] can be applied in (2) to extract the POD modes from the collected data. In all previous POD studies [29]-[44], material properties in the structure remain unchanged during the data collection. Therefore, each projected POD block represents a specific structure. If a material change in any location of the structure, a different POD model will be needed. In the MPV-POD approach for interconnects, a group of blocks including different metal/via routings with appropriate values of $\rho C$ and $k$ are performed in DNSs to embed effects of MPV into the POD modes. However, even if the generated MPV-POD modes include the MPV information, the POD parameters given in (7) and (8) still need to be evaluated numerically before each POD simulation because the material in the structure for each simulation may be different. This imposes a time-consuming process for thermal prediction of a large structure. This is however not a serious problem for interconnects. Using the example of the generic block in Fig. 3, one

can pre-evaluate POD parameters from integrals in (7) and (8) for this block without the regions where metal lines/vias may run. The parameters for this *blank* generic block with only the dielectric and the silicon substrate, can be stored in the library. Once the routings are determined in each block of this generic group, it will take little time to add integrals in (7) and (8) of the unfilled regions by filling either metal or oxide in the vacant regions before POD thermal simulation.

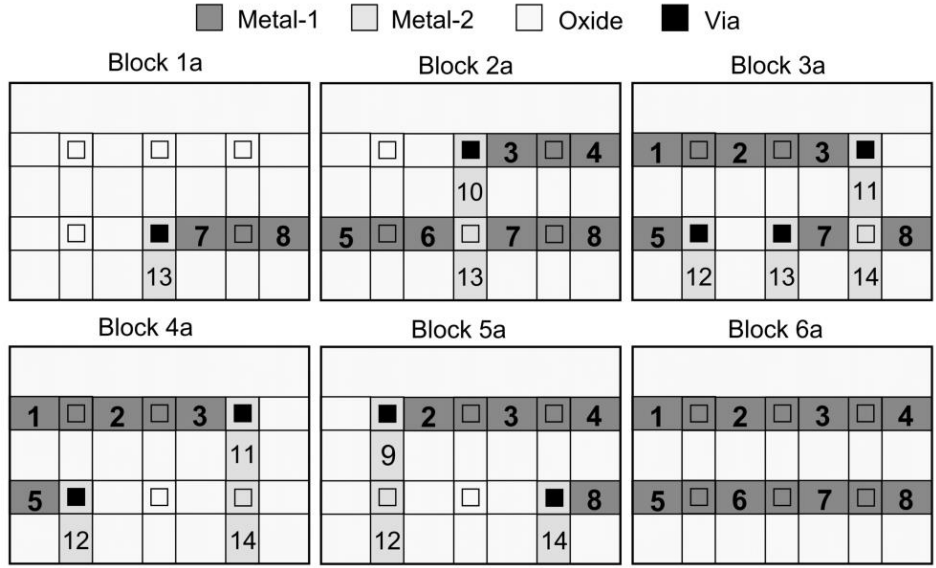

Fig. 4. Six interconnect blocks in Fig. 1. Blocks 1a-3a are on the top of the layout in Fig. 1 while Blocks 4a-6a are the mirror symmetric blocks of those on the bottom of Fig. 1. An open square at the via location indicates no via existence.

## 4. Illustration of MPV-POD Mode Generation

The generic interconnect block given in Fig. 3 is selected in this work to illustrate thermal data collection, mode generation and demonstrations for the MPV-POD methodology. The dimensions and material properties of the FinFET IC structure used in this study are adopted from an earlier investigation in [31].

To arrive at robust POD modes, the collected data from the selected interconnect blocks need to experience realistic variations of heating sources, BCs and material properties. To account for metal/oxide variations effectively along the routing paths, a simple guideline is used; namely each metal/via segment must appear a few times with different connections to other metal/via segments. For this test case, the 6 interconnect blocks given in Fig. 4 (adopted from the circuit in Fig. 1) are included to collect thermal data in DNSs. In addition, 6 more selected blocks with different metal/via routings shown in Fig. 5 are incorporated in the data collection. This ensures that each metal/via segment in these 12 blocks appears at least twice with connections to different metal/via segments. One can always include many more different connections for each segment to enhance the quality of the collected data and thus the robustness of the POD modes. This however will produce more data and may become computationally intensive in data collection, eigenvalue and mode generation, and calculations of POD model parameters from (7) and (8).

DNSs of each block in Figs. 4 and 5 need to perform to collect dynamic thermal data with spatial details, subjected to joule heating in the metal and BCs induced by the neighboring blocks. To account for heat fluxes across boundaries of each block appropriately, each of the 12 selected block is embedded in a larger simulation domain with an extended length of 200nm beyond the selected block on each side, such as the diagram shown in Fig. 6 for Block 2a. Any metal line reaching a boundary of the selected block is extended to the boundary of the simulation domain. The extension allows us to apply a more realistic range of BCs on the selected block boundaries. For example, BCs on the oxide boundary of the selected block (inside the simulation domain in Fig. 6) near metal lines are affected heat flow along these lines outside the selected block. The metal line extension shown in Fig. 6 offers realistic heat flow across the interface between blocks shown in Fig. 1. These BC variations offer information for the generated POD modes to learn enough variations to be able to accurately predict the heat flow in the metal wires even with different BCs. It should be pointed out that this simplified domain setting, although covering the major heat flux paths across the boundaries of the selected interconnect block (Block 2a), does not account for some BCs in the multi-block demonstration presented in Section 5.2. This is indeed partially responsible for the error observed in the multi-block demonstration. One can always include more simulations to accommodate more metal routings around the selected block to cover more variations of the BCs for the selected block that will enhance the quality of the collected data. This will again increase the computing resources needed for the POD mode training and model parameter calculations.

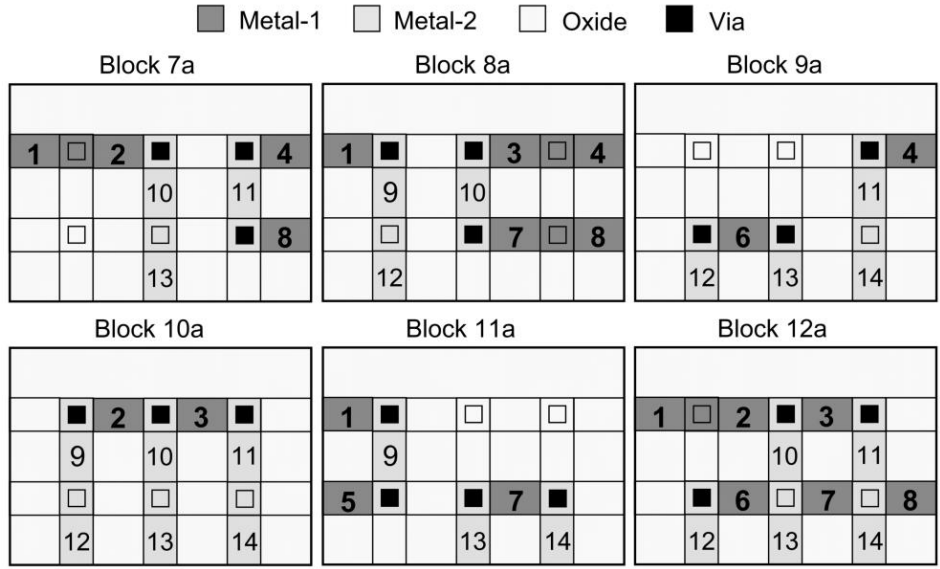

Fig. 5. Six additional blocks used in DNSs for thermal data collection to generate POD modes for the generic interconnect block given in Fig. 3.

Power density induced by Joule heating in (9) along metal lines results from current density. To apply more realistic dynamic current density, circuit simulation of the FinFET IC in Fig. 1 is performed in Spice. As shown in the circuit of Fig. 1, a 4 GHz voltage clock is applied to one of the inputs of each gate while a random digital voltage sequence is applied to the second input of each gate ($v_{i1}$ or $v_{i2}$). Dynamic current along each metal line is extracted from the Spice simulation and the power density is then implemented in the DNS of the simulation domain in Fig. 6. Even though the interconnect routings may be different between Figs. 1 and 6, the variations of the current and power density cover a range of metal BCs and joule heating for the generate POD modes to learn. The heat flux applied to each metal boundary of the simulation domain is consistent with the power density applied to its metal line. Ambient temperature is applied to the bottom of the substrate and the top boundary is adiabatic. On other boundary surfaces (oxide boundaries sown in Fig. 6), dynamic heat fluxes across the boundaries between the interconnect blocks extracted from the DNSs of the FinFET circuit in Fig. 1, together with some random variations, are applied. This offers more realistic variations of the boundary fluxes affected by neighboring interconnect blocks.

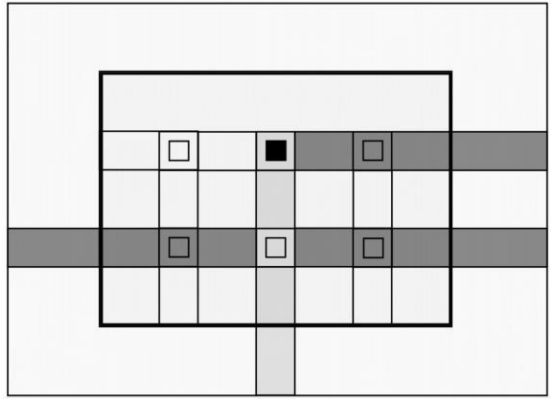

Fig. 6. Block 2a placed in a larger dielectric domain for DNSs for thermal data collection. The extended length of the domain in each direction is 200 nm.

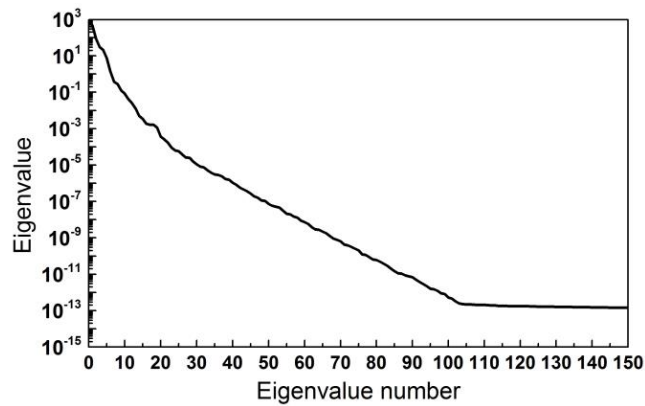

Fig. 7. Eigenvalue of collected thermal data from the 12 selected blocks.

DNS of each selected block was performed over 10 clock periods with 50 time steps in each period. Thermal data are collected at each time step. It should be noted that simulation settings for data collection (in turn for POD mode training) are not unique. Thermal data collected from DNSs of the selected blocks should as much as possible encompass a range of parametric variations which the generic blocks will experience in realistic operation, including variations of heat sources, BCs and metal/oxide along the interconnect routing paths.

Thermal data of the 12 interconnect blocks collected from the DNSs are combined together to generate eigenvalues and POD modes from (2) using the method of snapshots [31], [54]. The generic block in Fig. 3 is thus projected onto a POD space represented by these POD modes. The eigenvalue $\lambda_i$ represents the mean squared temperature variations captured by the $i$-th mode $\varphi_i$ and thus reveals the importance of the mode. The eigenvalue for the collected data shown in Fig. 7 decreases by more than 600 times from the first POD mode to the 6th mode and more than 4 orders of magnitude at 10th mode. The eigenvalue curve becomes nearly flat beyond the 102nd mode due to the machine precision.

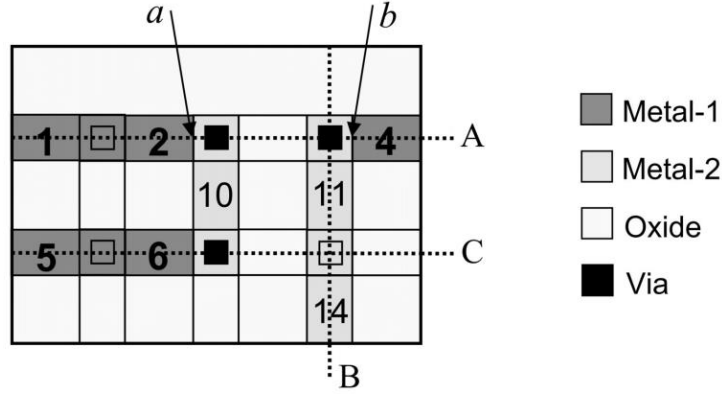

Fig. 8. Interconnect test block for the single-block demonstration. Points *a* and *b* are in the center of M1 lines at distances of 0.24µm and 0.45µm from the left boundary, respectively. Lines A and C run through the ceters of M1 lines and Line B through the M2-line center for temperature plots given in Fig. 10.

## 5. Demonstration of the MPV-POD Methodology

### 5.1. Single-block demonstration

With the POD modes generated from the thermal data collected from the DSNs of the 12 selected blocks given in Figs. 4 and 5, the MPV-POD approach is first demonstrated below in a single-block domain shown in Fig. 8. To test how the modes via the training accommodate different metal patterns, two distinct metal/via paths are included in the test block. The left metal/via path in the test block of Fig. 8 is different from any of the 12 selected blocks, and the right path is the same as the right metal/via path in Block 9a. Different random pulses of current density and different BCs are applied to the test block except the top adiabatic boundary and the ambient bottom boundary. To have meaningful comparison with DNS, current density, BCs and metal/via routings are identical in the simulations between these two approaches.

Dynamic temperatures at Points *a* and *b* specified in Fig. 8 are illustrated in Fig. 9 with the MPV-POD approach compared against the DNS. Because Point *b* is in a metal/via path identical to the right metal/via path in Block 9a, only 3 modes in the POD model are needed to reach an excellent agreement with DNS, as shown in Fig. 9(c) and 9(d). On the other hand, errors with 3 modes shown in Fig. 9(a) and 9(b) are relatively large at Point *a*. With 6 modes in POD, both approaches are in a very good agreement, where the POD model reaches a maximum deviation of 2.47% from DNS near a peak temperature at 0.675 ns.

Spatial temperature profiles at 0.675s in the test block are also shown in Fig. 10. Similar to Fig. 9, temperatures derived from the 3-mode MPV-POD model along Line B in M2 Segments 11 and 14 (Fig. 10(b)) and along Line A in M1 Segment 4 (beyond 4 µm in Fig. 10(a)) are in excellent agreement with that from the DNS. This is because the metal/via path from M1 Segment 4 to M2 Segments 11 and 14 (see Fig. 8) is identical to one of paths in Block 9a used in data collection. Along Lines A and C on the other metal/via path, the error derived from the MPV-POD model with 3 modes are however slightly large. With 6 modes in POD, a maximum error near 3% along Lines A and C is observed compared to the DNS.

The LS errors of the MPV-POD approach for different number of modes over the whole simulation domain and time are given in Table 1. For this single-block case, the error decreases as the number of modes increases and remains near 1.81% beyond 8 modes. With 6 or more modes, an LS error smaller than 2% can be achieved. The demonstration reveals that even for an interconnect routing not included in POD mode training, the MPV-POD model with a small number modes is still able to offer a very accurate prediction.

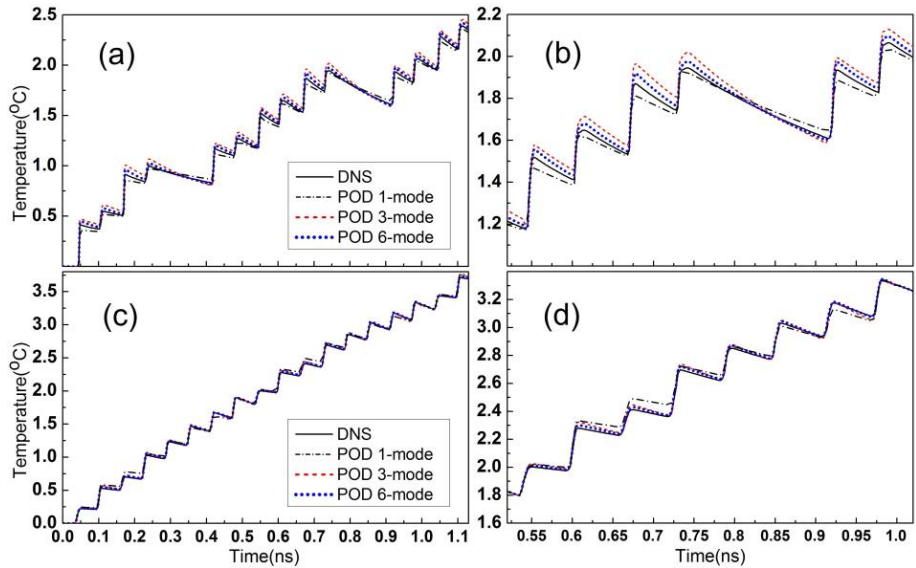

Fig. 9. Dynamic temperatures in M1 (a) at Point *a* specified in Fig. 8 with a more detailed illustration in (b) from 0.52ns-1.025ns, and (c) at Point *b* specified in Fig. 8 with a more detailed illustration in (d).

Table 1. LS error of single-block MPV-POD models

| No. of modes | 1 | 3 | 6 | 8 | 10 | 11 | 12 |
|---|---|---|---|---|---|---|---|
| LS error (%) | 6.82 | 3.14 | 1.97 | 1.82 | 1.81 | 1.8 | 1.81 |

*5.2. Multi-block demonstration in an integrated circuit*

To verify the multi-block MPV-POD methodology under a more realistic operation, the multi-block approach is implemented in a FinFET IC structure shown in Fig. 11. The layout in Fig. 11 represents the same circuit in Fig. 1 with NAND2 Gates B and C swapped in the layout. The POD modes for the generic interconnect block in Fig. 3 and its mirror symmetric block are applied to all the six interconnect blocks in Fig. 11 including Blocks 1b-3b and 4b′-6b′. In addition to the 6 interconnect blocks represented by the generic MPV-POD block, the POD thermal simulation of the entire IC structure also includes 3 additional NAND2 blocks described by one set of POD modes (without MPV effects) developed in [31]. The POD simulation is thus performed in a 9-block domain that is projected onto a POD space described by only 2 sets of POD modes, one set for the generic interconnect block and the other for the NAND2 block. The same number of modes are used for each of the 9 projected blocks in the demonstration. In the demonstration, a 4 GHz voltage clock is applied to the 3-NAND2 circuit. Different random digital voltages are applied to $v_{i1}$ and $v_{i2}$ in Spice simulation to estimate the power densities at device junctions and along each metal lines. These power densities are then implemented in both the POD simulation and DNS. In the simulation domain, the bottom of the substrate is fixed at ambient and all other boundaries are adiabatic except for the metal boundaries where surface power densities are applied based on the power evaluated from the Spice simulation.

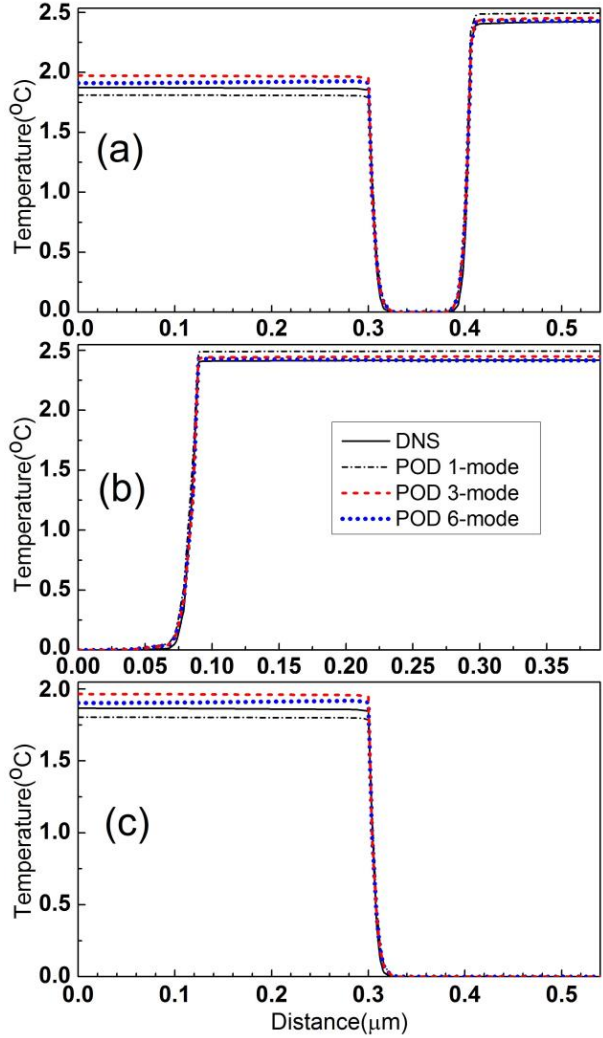

Fig. 10. Temperature profiles at 0.675 ns along (a) Line A in Fig. 8 through M1, (b) Line B in M2 and (c) Line C through M1.

Blocks 2b, 3b and 5b′ in Fig. 11 are different from any trained block (or their symmetric blocks) given in Figs. 4 and 5 while Blocks 1b, 4b and 6b are identical to Blocks 1a, 4a and 5a in Fig. 4, respectively. It should be noted that, even with Blocks 1b, 4b and 6b identical to 3 of the blocks in the data collection, the simulation settings for the 6 interconnect blocks in Fig. 11 are very different from those in the training. Simulation domains for the selected 12 blocks in the training process are similar to the structure given in Fig. 6 with the BCs that only emulate the effects induced by neighboring interconnect blocks. Also, there was no adiabatic boundary in these trained blocks. On the other hand, in the demonstration there is at least one adiabatic boundary on the dielectric surface of each block, as shown in Fig. 11. Moreover, one of the boundaries of each block in the demonstration is neighbored by a NAND2 block instead of an interconnect block. The VDD/GND M1 line in the NAND2 circuit shown in Fig. 11 runs in parallel closely with one of the boundary surfaces of each interconnect blocks. These M1 lines impose entirely different BCs for the interconnect blocks from the trained blocks where no metal line runs in parallel near any boundary.

Our study of this multi-block domain has found that the multi-block POD approach becomes numerically unstable if the penalty number $N_\mu$ is below $N_{\mu,min}$ or above $N_{\mu,max}$, where $N_{\mu,min} \approx 7$ and $N_{\mu,max}$

≈ 40. In the demonstration, the value of $N_\mu$ used in [31] ($N_\mu = 20$) is applied first. Other values of $N_\mu$ between 7 and 40 are then used to analyze its influence on the interface thermal discontinuity and the accuracy. The POD thermal modeling approach without considering the MPV for the FinFET IC has been investigated in detail in a previous study [31]. In this study, we only present thermal solution derived from the MPV-POD models in the interconnect blocks compared against the DNS.

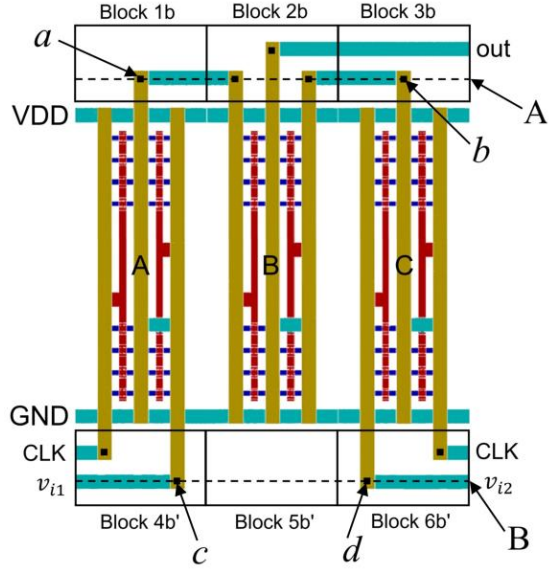

Fig. 11. Layout for the NAND circuit given in Fig. 1. This layput is however different from that in Fig. 1.

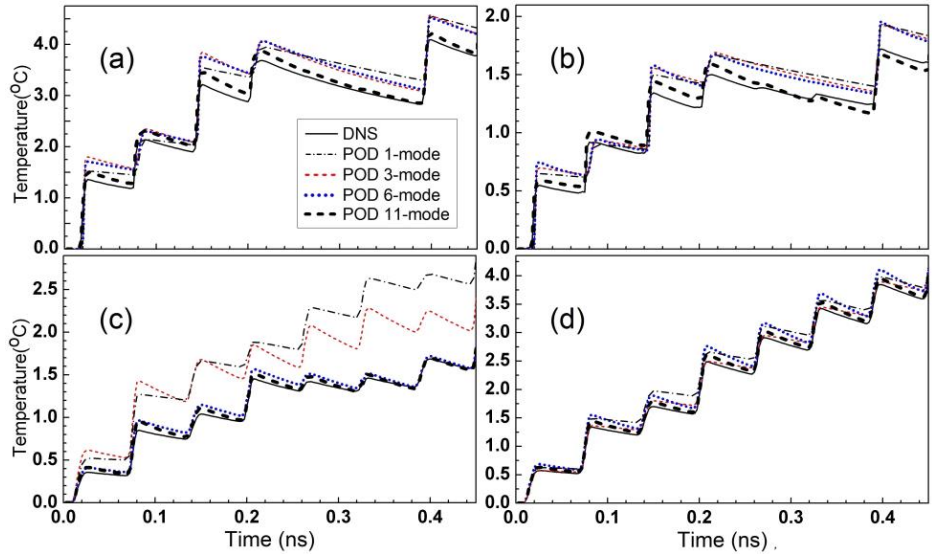

Fig. 12. Dynamic temperatures in M1 at (a) Point $a$, (b) Point $b$, (c) Point $c$ and (d) Point $d$ shown in Fig. 11 with $N_\mu = 20$ in the multi-block POD models.

Dynamic temperatures at Points $a$, $b$, $c$ and $d$ in M1 labeled in Fig. 11 under the vias are illustrated in Fig. 12. Even though Points $a$, $c$ and $d$ are located in the blocks identical to some selected trained blocks, due to inadequate boundary settings in the training, accuracy of dynamic temperatures in Fig. 12 ($N_\mu = 20$) for the multi-block MPV-POD model with 6 modes at these locations is not as good as those at Points $a$ and *b* in Fig. 9 for the single-block case. It needs 10 or 11 modes for the multi-block model to reach good accuracy. With 11 modes, maximum errors near 2.5%-4.4% around the peak temperatures at Points *a-d* are observed in Fig. 12. The LS error shown in Table 2 with $N_\mu = 20$ is near 3% with 11 modes and slightly reduces to 2.67% with 12 modes, that is still reasonably small considering that at least 2 BCs for each block are very different from the training settings, one with the adiabatic condition and the other induced by the VDD/GND M1 lines. Also, Point *b* is located in a block with very different metal/via routings from the trained ones. With very different BCs and metal/via routings in the training, the MPV-POD modes are somehow able to predict temperature profiles with good accuracy in the multi-block structure.

Table 2. LS error (%) of multi-block POD models

| No. of modes | 1 | 3 | 6 | 8 | 10 | 11 | 12 |
|---|---|---|---|---|---|---|---|
| $N_\mu = 7.3$ | 9.87 | 7.51 | 4.68 | 4.25 | 3.21 | 3.01 | 2.67 |
| $N_\mu = 20$ | 10.40 | 7.87 | 4.95 | 4.28 | 3.23 | 3.02 | 2.67 |
| $N_\mu = 37.2$ | 11.55 | 8.47 | 6.93 | 6.76 | 5.9 | 5.43 | 5.19 |

With only 3 modes, it is interesting to observe very accurate solution at Point *d* but poor accuracy at other locations. Also, much better accuracy is observed for the 6-mode MPV-POD model at Points *c* and *d* than at Points *a* and *b*. It should be noted that the POD process in (1) only optimizes the LS error instead of local errors. The LS error shown in Table 2 with $N_\mu = 20$ for the multi-block MPV-POD simulation reduces gradually, as more modes are used, and it reaches 2.67% with 12 POD modes. Even with more modes included in the multi-block case, its LS error only slightly reduces and is always greater than that of the single-block case. This is because accuracy of the multi-block case is limited by the data quality as a result of serious discrepancies of the BCs between the demonstration and the training. The other reason why more modes are needed for the multi-block MPV-POD approach to reach good accuracy is because of the boundary thermal discontinuities at block interfaces. Owing to the truncation of the solution given in (4), it is impossible to satisfy both the continuities of temperature and heat flux on any boundary. The interior penalty DG method [46], [47] is thus to enforce the flux continuity but allows a small temperature discontinuity (weak Dirichlet BCs) at the interface in an average sense between any 2 neighboring blocks.

The temperature distributions at 0.215ns along Lines A and B through the center of M1 lines (see Fig. 11) are illustrated in Fig. 13. Overall the MPV-POD model with more modes offers more accurate thermal solution; however the 3-mode model actually leads to a better accuracy than the 11 mode model in Block 6b′ along Line B, as shown in Fig. 13(b) and at Point *d* in Fig. 12. Along the metal lines in Line A, temperature discontinuities derived from the POD models with a small number of modes appear clearly across interfaces between neighboring blocks. With 3 modes, a discontinuity higher than 5% of the interface temperature is observed at each interface, as detailed in Figs. 14(a) and 14(b). Using 8 or 9 modes, the discontinuity is effectively suppressed to 2.2% or 1.5%, respectively, and it is successfully suppressed with 10 or more modes. Except for the one mode model, in general as the discontinuity is suppressed with more modes in the MPV-POD model, Table 2 shows that the LS error is reduced. Effects of the discontinuity is minimum along Line B because both interfaces are located in oxide.

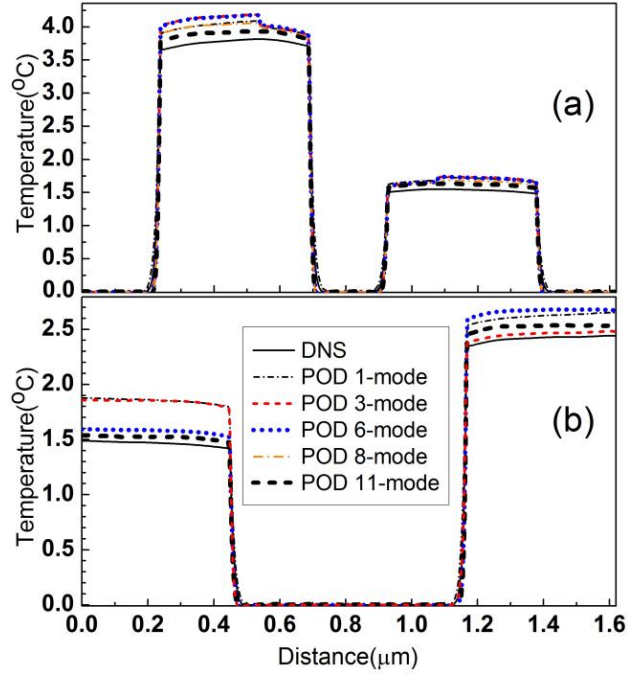

Fig. 13. Temperature profiles at 0.215 ns in Fig. 12 along (a) Line A and (b) Line B shown in Fig. 11 through the center of M1 line. In the multi-block POD models, $N_\mu = 20$.

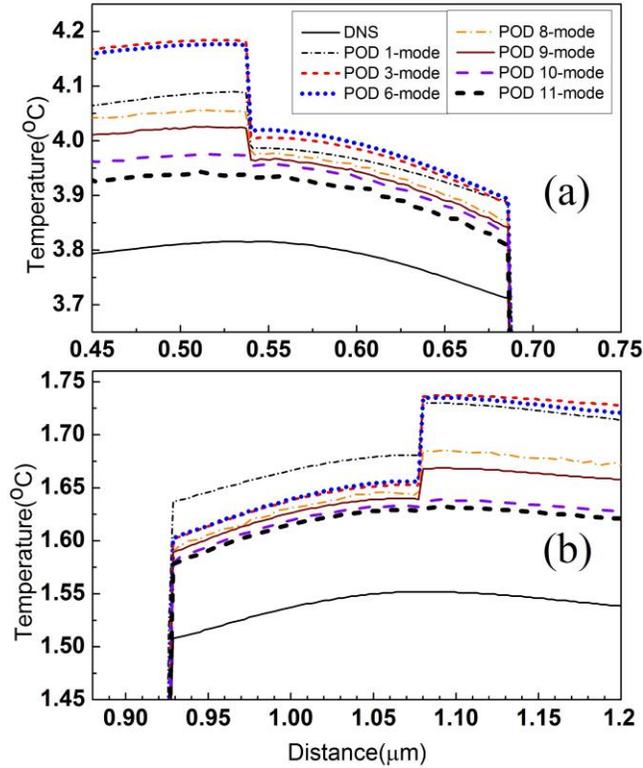

Fig. 14. Close-up views of the discontinuities on (a) the left and (b) the right of the temperature profiles in Fig. 13(a). In the multi-block POD models, $N_\mu = 20$.

To understand how $N_\mu$ influences the interface thermal discontinuities, different penalty numbers are applied in the POD simulations of the 9-block structure. Temperature profiles near the same interfaces in Fig. 14 are illustrated in Fig. 15 derived from the 8-mode and 11-mode POD models with $N_\mu$ slightly above $N_{\mu,min}$, slightly below $N_{\mu,max}$ and $N_\mu = 20$. Fig. 15 shows that the temperature discontinuity is suppressed and the accuracy of the POD prediction is improved as $N_\mu$ decreases from 37.2 to 7.3 for both 8-mode and 11-mode MPV-POD models. When $N_\mu = 37.2$, even with 11 modes in the MPV-POD model, discontinuities are still observed at bot interfaces in Figs. 15(a) and 15(b). As $N_\mu$ decreases to 20 or 7.3, the discontinuities are successfully removed. In addition, use of smaller values of $N_\mu$ leads to a better agreement with the DNS. As $N_\mu$ decreases from 37.2, 20 to 7.3, the improvement in the LS error for different number of modes is clearly observed in Table 2. A reduction of 1.5%-2.7% in the LS error can be achieved as $N_\mu$ changes from 37.2 to 7.3 with 6 or more modes. However, with 10 or more modes, use of $N_\mu$ equal to or below 20 is not able to further reduce the LS error and the LS errors for both $N_\mu = 7.3$ and 20 reach 2.67% with 12 modes even though the local errors near the interfaces shown in Figs. 15(a) and 15(b) reduce with a smaller $N_\mu$. It is believed that the LS error in this case is limited by the quality of the collected thermal data in the training of the generic block due to inadequate BCs, as discussed above.

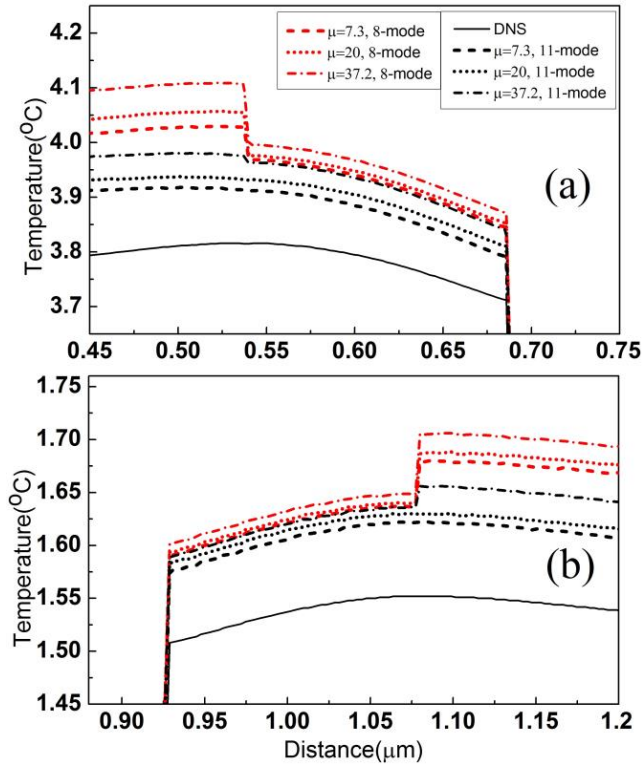

Fig. 15. Temperature profiles in (a) and (b) at the same interfaces as those in Figs. 14(a) and 14(b), respectively. In the multi-block POD models, the profiles with 8 and 11 modes are included for $N_\mu$ = 7.3, 20 and 37.2.

## 6. Discussions

It is found in this investigation on the single-block and multi-block interconnect structures that quality of the collected data, as a result of simulation settings in the training, is the key to determine the accuracy and robustness of the MPV-POD models. Appropriate settings of the simulation domain in the training of the POD modes are needed to cover enough variations of metal/via routings, power sources and BCs which the interconnect blocks will encounter in realistic operation.

In the single-block case, an MPV-POD model with only 3 modes offers excellent accuracy for dynamic temperature at Point *b* (between Segments 4 and 11 in Fig. 8) shown in Figs. 9(c) and (d) and for spatial temperature distributions shown in Figs. 10(a) and (b) along Segment 4 and Segments 11 and 14, respectively. This is because this metal/via routing is identical to a metal/via path in a selected trained block with perhaps slightly different BCs and power sources. Even with a metal/via routing in the test domain different from any selected trained block, the MPV-POD model with 6 modes was still able to offer a very good agreement with the DNS.

The simulation settings used to train the generic-block, although offering a good quality training for the single block case, does not provide good-quality data for the interconnect blocks embedded in the IC given in Fig. 11 due to serious inconsistent BCs between the demonstration and the training. Therefore, the POD modes of the generic block are not well-trained to adapt the BCs enforced by the IC operation in the multi-block case. As a result, the temporal/spatial temperature solution from the MPV-POD model in the multi-block case is not as good as that in the single-block demonstration. In addition, the multi-block approach suffers from inevitable thermal discontinuities across the interface between neighboring blocks. To further minimize the LS error in the multi-block structure, more POD modes are needed. It is also found that the value of the penalty number $N_\mu$ has a profound impact on the thermal discontinuities at block interfaces. An appropriate range for $N_\mu$ (in our case $7 < N_\mu < 40$) is needed to avoid numerical instability induced by the discontinuities, and a smaller $N_\mu$ within this range offers a smaller LS error. In the multi-block test case, even with inadequate BC training and different metal/via routings in 50% of the interconnect blocks, the MPV-POD modes of the generic interconnect block are able to offer a good thermal prediction, compared to with the DNS, as displayed in Table 2, if a smaller $N_\mu$ within the appropriate range is used.

Applications of the MPV-POD methodology to the single-block and multi-block interconnect structure reveal interesting and encouraging findings. This work demonstrates that, in addition to the heat sources and BCs, variations of material thermal properties between metal and oxide in interconnects can be captured effectively by the POD modes as long as enough variations of metal/via routings are implemented in the data collection (or the training). It has been shown that a small number of MPV-POD modes are able to offer a very accurate prediction of spatial/temporal thermal solution in interconnect blocks with routings that are different from the trained ones provided that the adequate BCs are included in the training. Even with inadequate BCs in the training, the MPV-POD model with 10 or more modes still offer a good description for the thermal solution in the multi-block interconnects. Apparently, the trained generic block was able to intelligently capture variations of metal/via routings and BCs to reach an accurate prediction for new routings or substantially different BCs as long as the block was trained by a wide range of routing and BC variations.

## 7. Conclusions

Use of building blocks has been one of the major practices for more effective engineering design and simulation in many different fields. In order to implement the building-block concept to improve effectiveness of thermal simulation of interconnects, MPV is proposed in POD to capture thermal effects of metal/via routings embedded in a dielectric structure. With this approach, some generic building blocks may be selected for a group of interconnect structure with similar metal/via routing features. Each selected generic block can be projected onto a POD space represented by a finite set of POD modes that are trained to capture variations of material properties, power sources and BCs. These generic blocks can then be glued together to construct a large interconnect structure. In this study, the interconnects used to wire the standard cells for a FinFET logic IC in Fig. 1 are partitioned into standard-size blocks modeled by a single generic block shown in Fig. 3. The selected generic block was trained to generate its POD modes and model parameters and then applied to demonstrate the accuracy and robustness of the MPV-POD approach in a single interconnect block and a multi-block IC structure.

The investigation reveals the importance of the quality of thermal data used to generate POD modes and parameters for the developed MPV-POD methodology. However, even with inadequate BCs accounted for in the training (that offers insufficient data quality), the MPV-POD model of the generic block is still able to offer a good prediction of the spatial/dynamic thermal simulation in the multi-block interconnect structure if more modes are used. It is also shown that the accuracy of the multi-block MPV-POD model is partially limited by the thermal discontinuities at block interfaces unless more modes are used. To the best of our knowledge, this study presents the first POD modeling approach with the MPV. It has been shown that it is possible to effectively account for thermal effects of the MPV induced by different metal/via routings in the POD modes for thermal simulation of interconnects. However, to derive robust MPV-POD models, in addition to a wide range of metal/via routings needed in the training of POD modes, BCs implemented in the training need to be close to those in realistic operation. To further improve the multi-block MPV-POD models, perhaps alternative schemes for more effectively minimizing the interface discontinuity need to be investigated.

## Acknowledgements

This work was partially supported by the National Science Foundation under Grant No. ECCS-2003307.

## Appendix: Brief Overview of Method of Snapshots

The autocorrelation tensor in (3) can be rewritten in terms of the average of the sample data sets over a number of snapshots (observations),

$$R(\vec{x},\vec{x}') = \frac{1}{N_t}\sum_{j=1}^{N_t} T(\vec{x},t_j)T(\vec{x}',t_j) , \tag{A1}$$

and thus (2) becomes

$$\frac{1}{N_t}\sum_{j=1}^{N_t} T(\vec{x},t_j)\int_\Omega T(\vec{x}',t_j)\varphi(\vec{x}')d\vec{x}' = \lambda\varphi(\vec{x}) , \tag{A2}$$

where $N_t$ is the total number of observations in time. In a static problem, the observations are performed at different heat source strengths and/or different BC's.

Let us now define the projection of the $j$th sample data set onto the POD space given by the integral in (A2) as

$$u_j = \int_\Omega T(\vec{x}',t_j)\varphi(\vec{x}')d\vec{x}' , \tag{A3}$$

and (A2) can then rewritten as

$$\frac{1}{N_t}\sum_{j=1}^{N_t} T(\vec{x},t_j)u_j = \lambda\varphi(\vec{x}). \tag{A4}$$

After multiplying both sides of (A4) by $T(\vec{x},t_i)$ and performing an integral on each side over the entire domain, the follow equation is obtained

$$\frac{1}{N_t}\sum_{j=1}^{N_t} u_j \int_\Omega T(\vec{x},t_i)T(\vec{x},t_j)d\vec{x} = \lambda \int_\Omega T(\vec{x},t_i)\varphi(\vec{x})d\vec{x}, \tag{A5}$$

which can be expressed as a matrix equation for a different eigenvalue problem,

$$\begin{bmatrix} A_{11} & \cdots & A_{1j} & \cdots & A_{1N_t} \\ \vdots & \ddots & \vdots & \iddots & \vdots \\ A_{i1} & \cdots & A_{ij} & \cdots & A_{iN_t} \\ \vdots & \iddots & \vdots & \ddots & \vdots \\ A_{N_t 1} & \cdots & A_{N_t j} & \cdots & A_{N_t N_t} \end{bmatrix} \begin{bmatrix} u_1 \\ \vdots \\ u_j \\ \vdots \\ u_{N_t} \end{bmatrix} = \lambda \begin{bmatrix} u_1 \\ \vdots \\ u_j \\ \vdots \\ u_{N_t} \end{bmatrix}, \tag{A6}$$

with $u_j$ given in (A3) and

$$A_{ij} = \frac{1}{N_t}\int_\Omega T(\vec{x},t_i)T(\vec{x},t_j)d\vec{x}. \tag{A7}$$

Once the eigenvectors in (A6) are determined, the POD modes at can be recovered using (A4) as a linear combination of observations,

$$\varphi(\vec{x}) = \frac{1}{N_t \lambda}\sum_{j=1}^{N_t} T(\vec{x},t_j)u_j. \tag{A8}$$

The eigenvalues derived from (A6) and the modes estimated (A8) have been shown in (A1)-(A8) to be identical to the first $N_t$ eigenvalues and POD modes, respectively, given in (2). For a large-scale domain with fine resolution, (2) represents an eigenvalue problem with an unmanageably large size, compared to the relatively small matrix size offered by (A6).